\documentclass[ALICE,manyauthors]{cernphprep}

\usepackage[comma,square,numbers,sort&compress]{natbib}
\usepackage{hyperref}
\usepackage{lineno}
\usepackage{ulem}
\usepackage{booktabs}
\usepackage{topcapt}

\newif\iflatexdiff
\latexdifftrue   
\usepackage{color}

\newcommand{\mev}          {\ensuremath{\mathrm{MeV}}}

\newcommand{\gevc}         {\ensuremath{\mathrm{GeV}/c}}

\newcommand{\gevcc}        {\ensuremath{\mathrm{GeV}/c^{2}}}

\newcommand{\pp}           {\text{pp}}

\newcommand{\PbPb}         {\mbox{Pb--Pb}}

\newcommand{\pPb}          {\mbox{p--Pb}}

\newcommand{\pt}           {\ensuremath{p_{\mathrm{T}}}}

\newcommand{\pT}           {\ensuremath{p_{\mathrm{T}}}}

\newcommand{\sqrtsnn}[1]   {\ensuremath{\sqrt{s_{\mathrm{NN}}}=#1~\mathrm{TeV}}}
\newcommand{\sqrts}[1]     {\ensuremath{\sqrt{s}=#1~\mathrm{TeV}}}

\newcommand{\com}[1]       {}

\newcommand{\RAA}          {\ensuremath{R_{\mathrm{AA}}}}
\newcommand{\RPBPB}          {\ensuremath{R_{\mathrm{PbPb}}}}
\newcommand{\RpPb}          {\ensuremath{R_{\mathrm{\it{p}\mathrm{Pb}}}}}

\iflatexdiff

\renewcommand{\xout}[1]    {\textcolor{red}{\sout{#1}}}
\newcommand{\ask}[1]       {\textcolor{magenta} {#1} }
\else

\renewcommand{\xout}[1]    {}
\newcommand{\ask}[1]       {}
\fi

\newcommand{\jpsi}         {\ensuremath{{\mathrm {J}}/\psi}}

\newcommand{\TAA} {\ensuremath{\langle T_{\mathrm{AA}} \rangle}}

\begin{document}%

\begin{titlepage}
\PHyear{2016}
\PHnumber{235}      
\PHdate{21 Sep}  
%

\title{Measurement of the production of high-\pt\ electrons from heavy-flavour hadron decays in \PbPb\ collisions at  $\mathbf{\sqrt{\it s_{\rm{NN}}}}$ = 2.76 TeV}
\ShortTitle{Heavy-flavour decay electrons in \PbPb\ collisions at \sqrtsnn{2.76}}   

\Collaboration{ALICE Collaboration\thanks{See Appendix~\ref{app:collab} for the list of collaboration members}}
\ShortAuthor{ALICE Collaboration} 

\begin{abstract}
Electrons from heavy-flavour hadron decays (charm and beauty) were measured with the ALICE detector in \PbPb\ collisions at a centre-of-mass of energy \sqrtsnn{2.76}.
The transverse momentum ($\pt$) differential production yields at mid-rapidity were used to calculate the nuclear modification factor \RAA in the interval $3<\pt< 18$ GeV/$c$.
The \RAA\ shows a strong suppression compared to binary scaling of pp collisions at the same energy (up to a factor of 4) in the 10\% most central \PbPb\ collisions.
There is a centrality trend of suppression, and a weaker suppression (down to a factor of 2) in semi-peripheral (50--80\%) collisions is observed.
The suppression of electrons in this broad $p_{\rm T}$ interval indicates that both charm and beauty quarks lose energy when they traverse the hot medium formed in Pb--Pb collisions at LHC.
\end{abstract}

\end{titlepage}
\setcounter{page}{2}

\section{Introduction}

High-energy heavy-ion collisions provide a unique opportunity to study the properties of the hot and dense strongly-interacting system composed of deconfined quarks and gluons -- the quark-gluon plasma (QGP).
The formation of a QGP is predicted by lattice QCD calculations \cite{Satz:2000bn,Bass:1998vz,Shuryak:1984nq,Cleymans:1985wb}. A crossover transition from hadronic matter at zero baryochemical potential is expected to take place once the system temperature reaches values above $T$$\approx$ 155 MeV and/or the energy density above $\epsilon$ $\approx$ 0.5 GeV/fm$^3$ \cite{Borsanyi:2010cj,Bhattacharya:2014ara}.
To characterize the physical properties of this short-lived QGP (lifetime of about 10 fm/$c$ \cite{Aamodt:2010jj}) experimental studies use auto-generated probes, such as high-energy partons created early in the collision, thermally emitted photons, and particle correlations sensitive to the collective expansion and the dynamics of the system.

In particular, the interaction of high-\pt\ partons with the QGP, leading to modifications of the internal jet structure (jet quenching), was first proposed in \cite{Bjorken:1982tu} and is studied as a sensitive probe of the medium properties \cite{Burke:2013yra}. Jet quenching was first observed experimentally via the strong suppression of high transverse momentum particle production in heavy-ion collisions at the Relativistic Heavy Ion Collider (RHIC) \cite{Adams:2005dq,Adcox:2004mh,Arsene:2004fa,Back:2004je}.
Similar observations have since been reported by the Large Hadron Collider (LHC) experiments at collision energies larger by one order of magnitude with hadrons \cite{Aamodt:2010jd, Aamodt:2011vg, CMS:2012aa} and extended to fully reconstructed jets \cite{Aad:2010bu, Chatrchyan:2011sx, Adam:2015ewa}.

Heavy flavours (charm and beauty) are sensitive tools for studies of the in-medium parton energy loss, providing qualitatively different sensitivity to the medium properties as compared to gluon or light-quark induced jets \cite{Andronic:2015hfov,Prino:2016cni}. The production of heavy quarks is well understood in terms of the perturbative QCD (pQCD) formalism. Good agreement between the theoretical calculations and measurements of various heavy-flavour particle production cross sections in hadronic collisions is established over a wide range of centre-of-mass energies from RHIC \cite{Adare:2010de, Adams:2004fc, Adamczyk:2012af}, through the Tevatron \cite{Acosta:2003ax, Abazov:2008rc, Abazov:2008af} to the LHC \cite{Chatrchyan:2012dk,ATLAS:2011ac,Abelev:2014gla,Abelev:2012qh,Abelev:2012tca}.

Interactions between partons and the medium can occur via both inelastic (radiative parton energy loss) \cite{Gyulassy:1990bh,Wang:1994fx,Baier:1996sk} and elastic (collisional energy loss) \cite{Thoma1991491,PhysRevD.44.1298,PhysRevD.44.R2625,Majumder:2008zg} processes that depend on the parton type and the properties of the medium.
The interactions with the medium modify the radiation pattern of the shower by inducing longitudinal drag (and associated longitudinal diffusion), transverse diffusion, and enhanced splitting of the propagating partons.
On average, for a given parton energy, gluons are expected to lose more energy than quarks due to the difference in the Casimir colour factor \cite{Agashe:2014kda} controlling the strength of the coupling to the coloured medium.
Moreover, the energy loss is predicted to depend on the mass of the quark \cite{Dokshitzer:2001zm,Armesto:2003jh,Djordjevic:2003zk,PhysRevLett.93.072301,Wicks:2007am}. In particular, for quarks with energies comparable to their mass the radiative energy loss is expected to be smaller than for more highly-energetic partons.
Consequently the relative role of elastic processes for heavy quarks is enhanced and the heavy quarks of moderate energies are expected to be more sensitive, as compared to light quarks, to the longitudinal drag and diffusion coefficients \cite{Majumder:2008zg} that are proportional to the inverse of the mass of the parton.
Moreover, as a result of multiple elastic collisions and possible in-medium resonant interactions within the hot matter, low-momentum heavy quarks could reach thermalisation in the medium \cite{vanHees:2005wb}.

The predicted hierarchy of energy loss $\Delta E_{\mathrm{g}} > \Delta E_{\mathrm{light-q}} > \Delta E_{\mathrm{charm}} > \Delta E_{\mathrm{beauty}}$ \cite{Dokshitzer:2001zm} motivated experimental studies of the suppression patterns of heavy-flavour hadrons and their decay products.
Up to now the in-medium energy loss of heavy flavours at the LHC has been studied via open charm measurements of prompt D mesons \cite{Adam:2015nna, Adam:2015sza}, heavy-flavour decay muon measurements at forward rapidity \cite{Abelev:2012qh}, non-prompt $\mathrm{J}/\psi$, and measurements of b-jet production \cite{Chatrchyan:2013exa}.
At RHIC the nuclear modification of heavy-flavour production has been studied via its semileptonic decays \cite{Adare:2013yxp, Abelev:2006db} and via measurements of D mesons \cite{Adamczyk:2014uip}.
The measurement that we report covers the electron (electron and positron) \pt\ interval 3--18 GeV/$c$, probing at high \pt\,
the in-medium interaction of b quarks with momentum of a few tens of GeV/$c$.

The modifications of particle yields are quantified using the nuclear modification factor \RAA. It is constructed by dividing the \pt-differential yield in nucleus-nucleus (AA) collisions, $\mathrm{d}N_{\mathrm{AA}}/\mathrm{d}p_{\mathrm T}$, by the cross section in \pp\ collisions, $\mathrm{d}\sigma_{\mathrm{pp}}/\mathrm{d}p_{\mathrm T}$, scaled by the average of the nuclear overlap function \TAA\ for the considered centrality class \cite{Abelev:2013qoq}
\begin{equation}
\RAA = \frac{\mathrm{d}N_{\mathrm{AA}}/\mathrm{d}p_{\mathrm{T}}}{\TAA \mathrm{d}\sigma_{\mathrm{pp}}/\mathrm{d}p_{\mathrm{T}}}.
\label{RAAeq}
\end{equation}
By construction, $R_{\textrm {AA}}$ is unity when no nuclear effects are present.
 \RAA\ values consistent with unity have been measured for colour neutral particles (direct photons, $\mathrm{W}$ and $\mathrm{Z}$ bosons) in \PbPb\ collisions at \sqrtsnn{2.76} \cite{Chatrchyan:2012vq,Chatrchyan:2012nt,Chatrchyan:2011ua,Aad:2013zb,Aad:2014bha} as well as for charged particles and heavy-flavour production in \pPb\ collisions at \sqrtsnn{5.02} \cite{Abelev:2014dsa, Abelev:2014hha, Adam:2015qda}.

This paper reports on the suppression ($\RAA < 1$) of electrons from semi-leptonic decays of charm and beauty hadrons measured at high-transverse momentum ($\pt>3$~\gevc) at mid-rapidity ($|y|<0.6$) in \PbPb\ collisions at \sqrtsnn{2.76} using the ALICE detector. The suppression is measured as a function of collision centrality and $p_{\rm{T}}$ in the interval $3<p_{\mathrm{T}}<18$ GeV/$c$. The next two sections of the paper define the experimental setup and the analysis details together with the discussion of systematic uncertainties on the measured electron spectra. The electron yields measured in bins of centrality defined as fractions of the total hadronic cross-section of \PbPb\ collisions are then presented. Finally the $p_{\textrm T}$--differential \RAA\ in the 0--10\%, 10--20\%, 20--30\%, 30--40\%, 40--50\% and 50--80\% centrality classes are presented and compared to the measurement of muons from heavy-flavour hadron decays at forward rapidities \cite{Abelev:2012qh} as well as to calculations of in-medium energy loss of heavy quarks.

\section{Apparatus, data sample and analysis }

\subsection{Detector setup}

The measurements were carried out using the ALICE detector at the LHC \cite{Evans:2008zzb} with Pb-ion beams at a centre-of-mass energy of \sqrtsnn{2.76}.
A complete description of the experimental setup and the performance of detectors can be found in \cite{Aamodt:2008zz,Abelev:2014ffa}.
Particle track reconstruction and particle identification were performed based on information from the Inner Tracking System (ITS), the Time Projection Chamber (TPC), and the Electromagnetic Calorimeter (EMCal), located inside a solenoid magnet, which generates a $0.5$ T field parallel to the beam direction.
The event centrality determination was based on the signals from the V0 detector, which is a set of scintillator arrays.
Moreover, the V0 detector together with the neutron Zero-Degree Calorimeters (ZN) was used for triggering and beam background rejection.

The ITS is composed of six cylindrical layers: two Silicon Pixel Detectors (SPD), two Silicon Drift Detectors (SDD), and two Silicon Strip Detectors (SSD). The SPD barrel consists of staves distributed in two layers around the beam pipe at radius of 3.9 cm and 7.6 cm, covering a length of 28.2 cm in the $z$ direction. The outermost layer of the ITS (SSD) is located 43 cm from the beam axis.

The TPC with a radial extent of 85--247 cm, enables charged particle tracking beyond the ITS and particle identification via the measurement of the particle specific ionisation energy loss within the  Ne-CO$_{2}$ gas mixture. The TPC provides up to 159 independent space points per particle track.

Charged particle tracks are reconstructed in the TPC from $p_{\textrm T}\approx$ 0.15 GeV/$c$, $|\eta| < 0.9$ and full azimuth. Using the ITS and TPC space points the particle momentum is determined from the combined track fit with a resolution of about 1\% at $1$~\gevc\ and about 3\% at $10$~\gevc\ \cite{Abelev:2014ffa}.

The front face of the EMCal is positioned at about 450~cm from the beam axis in the radial direction and the detector is approximately 110~cm deep.
The detector is a layered Pb-scintillator sampling calorimeter covering 107 degrees in azimuth and a pseudorapidity region $|\eta| < 0.7$. The calorimeter design incorporates on average a moderate active volume density that results in a compact detector of about 20 radiation lengths.

The V0 detector consists of two arrays of 32 scintillator tiles placed at distances $z = 3.4$ m (V0-A) and $z = -0.9$ m (V0-C) from the nominal interaction point. V0-A and V0-C cover the full azimuth, and pseudorapidity intervals of $2.8 < \eta < 5.1$ and $-3.7 < \eta < -1.7$, respectively. The detector was used for triggering and event centrality determination.

The ZN are two identical sets of forward hadronic calorimeters which are located on both sides relative to the interaction point at $z \approx$ 114~m.

\subsection{Event sample and trigger}

The data sample used for this analysis was collected in 2011 and it consists of $14 \cdot 10^{6}$ most central collisions (0--10\%) and $13 \cdot 10^{6}$ semi-central collisions (10--50\%) recorded with a minimum-bias trigger, and $3.2 \cdot 10^{6}$ collisions (0--90\%) triggered with the EMCal.
The minimum-bias trigger was a coincidence of signals from the V0-A and V0-C detectors. The timing resolution of the V0 system is better than 1 ns and it provides an efficient discrimination of the beam-beam collisions from the background events produced upstream of the experiment. Additional suppression of the background was provided by timing information from the ZDC.
The minimum-bias trigger included two trigger classes for most-central and semi-central collisions, which were selected online by applying thresholds on the V0 signal amplitudes.

The EMCal provides two hierarchically-configured trigger levels (Level-0 and Level-1). For this analysis the data were recorded with the L1 trigger in coincidence with the V0 minimum-bias trigger.
The trigger logic of the Level-1 trigger employed a sliding window algorithm of 4$\times$4 towers with a sliding step of 2 towers along either of the surface axes.
An event was rejected unless the energy summed within at least one set of the 16 adjacent towers was greater than a threshold.
Additionally, the trigger logic was configured to adjust the online threshold according to the event centrality estimated from the analogue sum of the V0 detector signals.
The threshold was adjusted such that the rejection rate was approximately constant as a function of the event centrality.
The thresholds varied from 7~GeV in 10\% most central events to 2~GeV in the most peripheral events.

The offline selection retained only events where the
coordinate of the reconstructed vertex along the beam direction was within \mbox{$\pm10$~cm} around the nominal interaction point.
The event vertex reconstruction is fully efficient for the event centralities considered in this analysis.

Collisions were classified into different centrality classes in terms of percentiles of the hadronic \PbPb\ cross section using the signal amplitudes in the V0 detector. The event centrality was related to
the nuclear overlap function $T_{{\rm{AA}}}$ via a Glauber model \cite{Miller:2007ri}.
Details on the centrality determination can be found in \cite{Abelev:2013qoq}.

\begin{figure}
\begin{center}
\includegraphics[width=0.92\textwidth,viewport=10 5 540 315,clip]{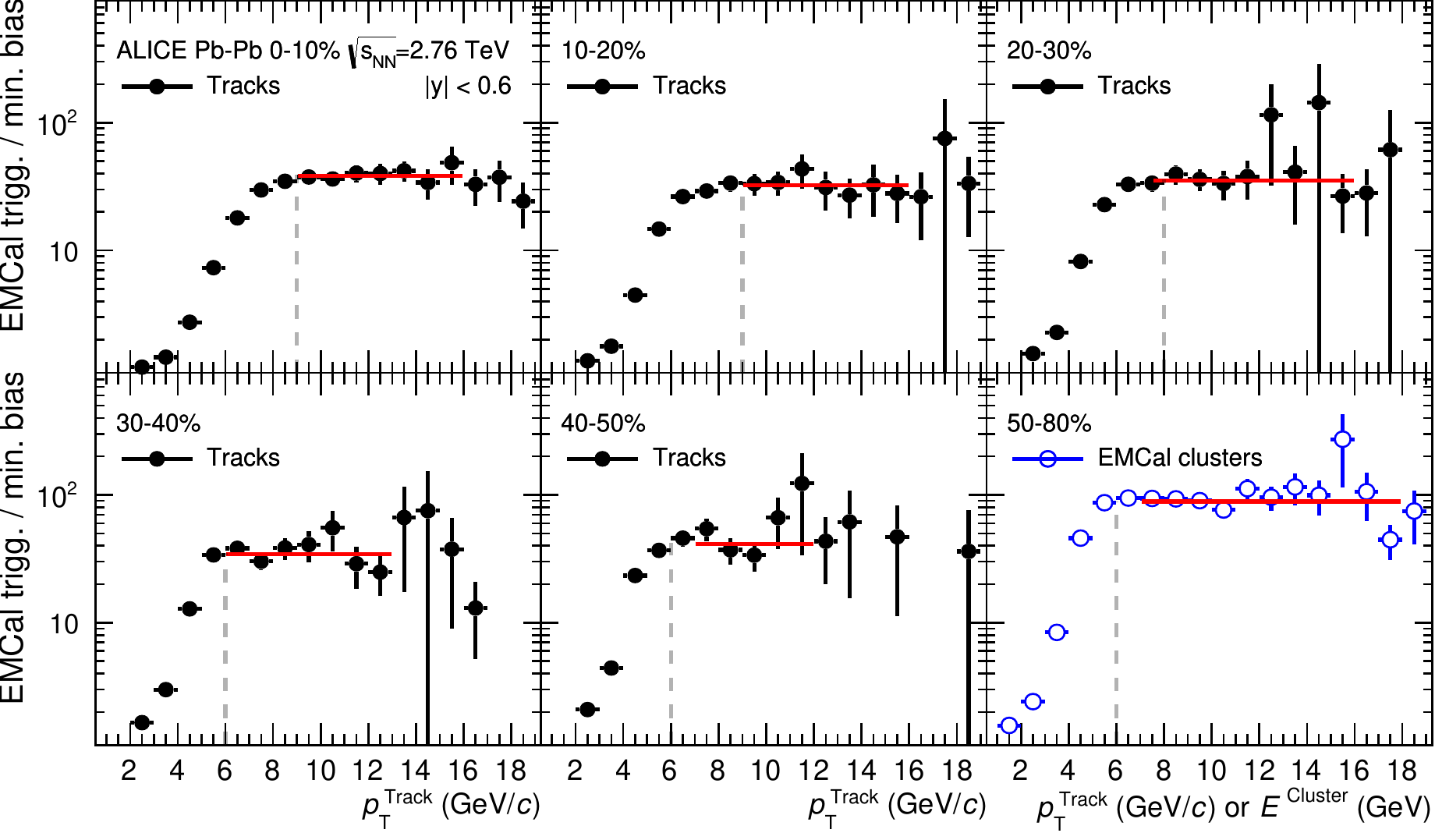}
\end{center}
\caption{Trigger turn-on curves: the ratio of inclusive electrons in EMCal triggered events to
  minimum-bias events as a function of associated track \pt\ in centrality bins between 0\% and 50\%. The lower right panel shows a similar ratio obtained with EMCal clusters for centrality 50--80\%. The \pt\ from which the spectra from the minimum-bias trigger to the EMCal trigger are used are indicated with black dashed lines. The scaling factors which were obtained by fits (red lines) are summarized in Table \ref{tab:rejectionFactors}.}
\label{fig:t-curve}
\end{figure}

To obtain the inclusive electron spectra utilizing minimum bias and EMCal triggers, in each centrality class, the per-event yield of electrons from the EMCal triggered sample was scaled to the minimum-bias yield by normalisation factors determined with a data-driven method.
Figure \ref{fig:t-curve} shows the ratio of \pt\--differential yields of the electron candidate tracks from the EMCal triggered sample to the minimum-bias trigger sample as a function of the track \pt.
The electron candidates were selected based on the ionisation energy loss in the TPC gas and the ratio of the EMCal cluster energy and the momentum of the particle track
(details of electron identification are given in the next section).
Because of the limited electron yield in the semi-peripheral event class (50--80\%) the correction for the trigger enhancement in that interval was obtained as the ratio of the energy distributions of EMCal clusters for the two trigger types (shown in panel (f) of Fig.~\ref{fig:t-curve}).
The inclusive \pt\ spectrum of electrons is formed by the electron spectrum from minimum-bias events below the trigger plateau (indicated by dashed lines in Fig.~\ref{fig:t-curve}) and the spectrum measured with only the EMCal trigger in the plateau region.
The difference in the shape of the curves in Fig. \ref{fig:t-curve} for $p_{\rm{T}}$ below the plateau is a consequence of the particle mixture contributing to the EMCal clusters and response of the EMCal to charged hadrons.
The scaling factors and the transition from the minimum-bias sample to the triggered sample were determined by fits with a constant to the high-\pt\ plateau regions.
The scaling factors for all centrality classes as well as the \pt\ at which the switch from the minimum-bias to the EMCal trigger spectra occurs are summarized in Table~\ref{tab:rejectionFactors}.
The uncertainty on the factors (also reported in Table~\ref{tab:rejectionFactors}) was obtained from the individual fits and therefore it is driven by the statistical uncertainty of the measured spectra.
The scaling factors within centralities 0--50\% were extracted using the electron tracks, whereas for centralities larger than 50\% the spectrum of EMCal clusters is used.
The relative difference in the scaling factors depending on whether electrons or clusters were selected was studied and shown to be below 8.5\%. This difference was included in the systematic uncertainty of the measurement for all centrality classes.
Table \ref{tab:rejectionFactors} corresponds to Fig. \ref{fig:t-curve}.

\begin{table}[t!]
   \centering
   \begin{tabular}{@{} m{2.5cm} m{2.5cm} m{2.5cm} @{}}
      \toprule
      Centrality & Scaling factor & Plateau above \pt\ (\gevc) \\
      \midrule
      0--10\%  & 38   $\pm$ 2.2 & 9 \\
      10--20\% & 32   $\pm$ 3.5 & 9 \\
      20--30\% & 35   $\pm$ 2.9 & 8 \\
      30--40\% & 34   $\pm$ 2.5 & 6 \\
      40--50\% & 41   $\pm$ 5.2 & 6 \\
      50--80\% & 89 $\pm$ 3.5 & 6 \\
      \bottomrule
   \end{tabular}
   \caption{Summary for centrality dependence of the EMCal trigger scaling factor. Middle column: trigger scaling factors (together with their absolute statistical uncertainty) extracted from the ratio of electrons (or EMCal cluster) \pt\ spectra in EMCal triggered and minimum-bias events. Right column: particle \pt\ at which the spectrum measured in minimum-bias events and EMCal triggered events are switched to form the inclusive electron \pt\ spectrum. See text for details.}
   \label{tab:rejectionFactors}
\end{table}

\subsection{Electron reconstruction}

For the reconstruction of electrons in this analysis tracks with a minimum of 100 out of 159 possible TPC space points were retained.
In addition, tracks were selected using their distance of closest approach (DCA) to the primary vertex.
Accepted tracks were within $|{\rm DCA}_{xy}| < 2.4$~cm in the transverse plane and $|{\rm DCA}_{z}| < 3.2$~cm along the beam axis.
Furthermore, the tracks were selected within a fiducial pseudorapidity acceptance of $|\eta| < 0.6$.
Each track was required to contain at least one point measured in the SPD and at least three hits out of the maximum of six in the ITS.
Moreover, the electron candidates were selected by applying a cut on the specific ionisation energy loss ($\mathrm{d}E/\mathrm{d}x$) within the TPC. The measured $\mathrm{d}E/\mathrm{d}x$ was required to be between $-1$ to 3$\sigma$, where $\sigma$ is $\mathrm{d}E/\mathrm{d}x$ resolution, from the expected mean of $\mathrm{d}E/\mathrm{d}x$ for electrons.
This selection is hereafter indicated as $-1<n_{\sigma}^{\mathrm{TPC}}<3$.
The tracks extrapolated to the sensitive volume of the EMCal were matched with a cluster if the cluster-track residual in azimuth and pseudorapidity was within a window of $|\Delta \varphi| < 0.05$ and $|\Delta \eta| < 0.05$.
Such matching criteria corresponds to an effective radius of about 6 times larger than the effective Moliere radius for EMCal, thus it is fully efficient for electron tracks with $\pt > 2~\gevc$.

\begin{figure}
\begin{center}
\includegraphics[width=1.0\textwidth,viewport=10 0 560 280,clip]{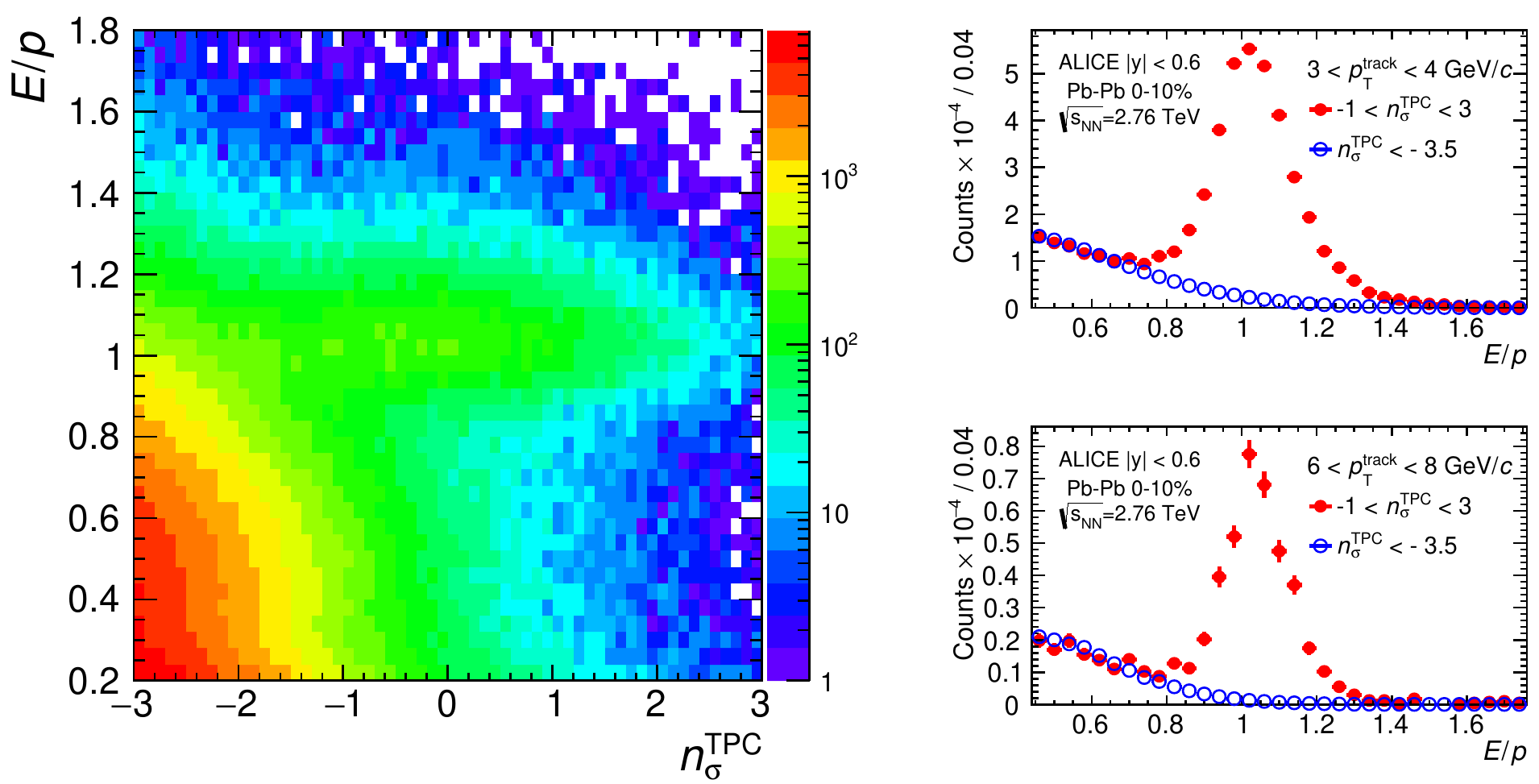}
\end{center}
\caption{{\it \bf Left:} The ratio of $E/p$ as a function of $n_{\sigma}^{\mathrm{TPC}}$ in 10\% most central Pb--Pb events ($p_{\textrm{T}}>$ 3 GeV/$c$), where $p$ is the charged particle momentum, $E$ is the matched EMCal cluster energy, and $\sigma^{\mathrm{TPC}}$ is the resolution on the energy loss in the TPC gas expected for electrons. {\it \bf Right:} $E/p$ for electrons in two transverse momentum ranges. The blue open symbols shows the hadron contamination -- an $E/p$ distribution for particles 3.5 $\sigma$ away from the mean of the true electron ${\rm TPC}$-d$E$/d$x$ distribution normalized to the electron $E/p$ at small values of the ratio (away from the electron signal).}
\label{fig:nsigmaeop}
\end{figure}

Additional hadron rejection used the combination of the energy deposited within EMCal and a cut on the electromagnetic shower shape \cite{Abelev:2014ffa,Abelev:2012sca}. Since the shower from an electron is fully contained and accurately measured by the EMCal, the ratio of the energy ($E$) measured by the EMCal and the momentum ($p$) for electron tracks is approximately unity ($E/p \approx 1$). The $E/p$ distribution is qualitatively different in the case of hadrons.
The $E/p$ as a function of the $n_{\sigma}^{\mathrm{TPC}}$ for charged particles matched with an EMCal cluster in 10\% most central events is shown in Fig. \ref{fig:nsigmaeop}.
From the primary tracks matched to an EMCal cluster the electron candidates were selected using a momentum independent cut of $0.9 < E/p < 1.3$. Furthermore, the shapes of the measured showers in the calorimeter can be characterized by the two eigenvalues ($\lambda_{0}$ and $\lambda_{1}$) of the covariance matrix built from the tower coordinates weighted by the logarithms of the tower energies.
These eigenvalues may be used to differentiate between incident particle species \cite{Abelev:2014ffa}.
A selection of $\lambda_{1}^{2} < 0.3$, corresponding to the shorter-axis of the shower shape projected onto the EMCal surface, was applied, because the characteristic electromagnetic shower of an electron is peaked at $\lambda_{1}^{2}$ of about 0.25 independent of the cluster energy.

The remaining hadron background in the electron sample was estimated with a data-driven approach and statistically subtracted from the sample. The shape of the residual hadron background in $E/p$ at the position of the electron peak was reconstructed using the $E/p$ distribution for hadron-dominated tracks selected with $n_{\sigma}^{\mathrm{TPC}} < -3.5$. The $E/p$ distribution of the hadrons was then normalized to match the distribution of the electron candidate in $0.4  < E/p < 0.7$ (away from the true electron peak). An example of the $E/p$ distributions together with the estimated hadron contamination for two transverse momentum intervals is shown in the right panel of Fig. \ref{fig:nsigmaeop}.
The hadron contamination is less than 5\% at $p_{T} < 10$ GeV/c in all centrality classes.
At high $p_{T}$, it is larger than 10\% with a maximum of about 15\% at $p_{T}$ = 18 GeV/c.

\begin{figure}
\begin{center}
\includegraphics[width=1.0\textwidth,viewport=5 20 550 260,clip]{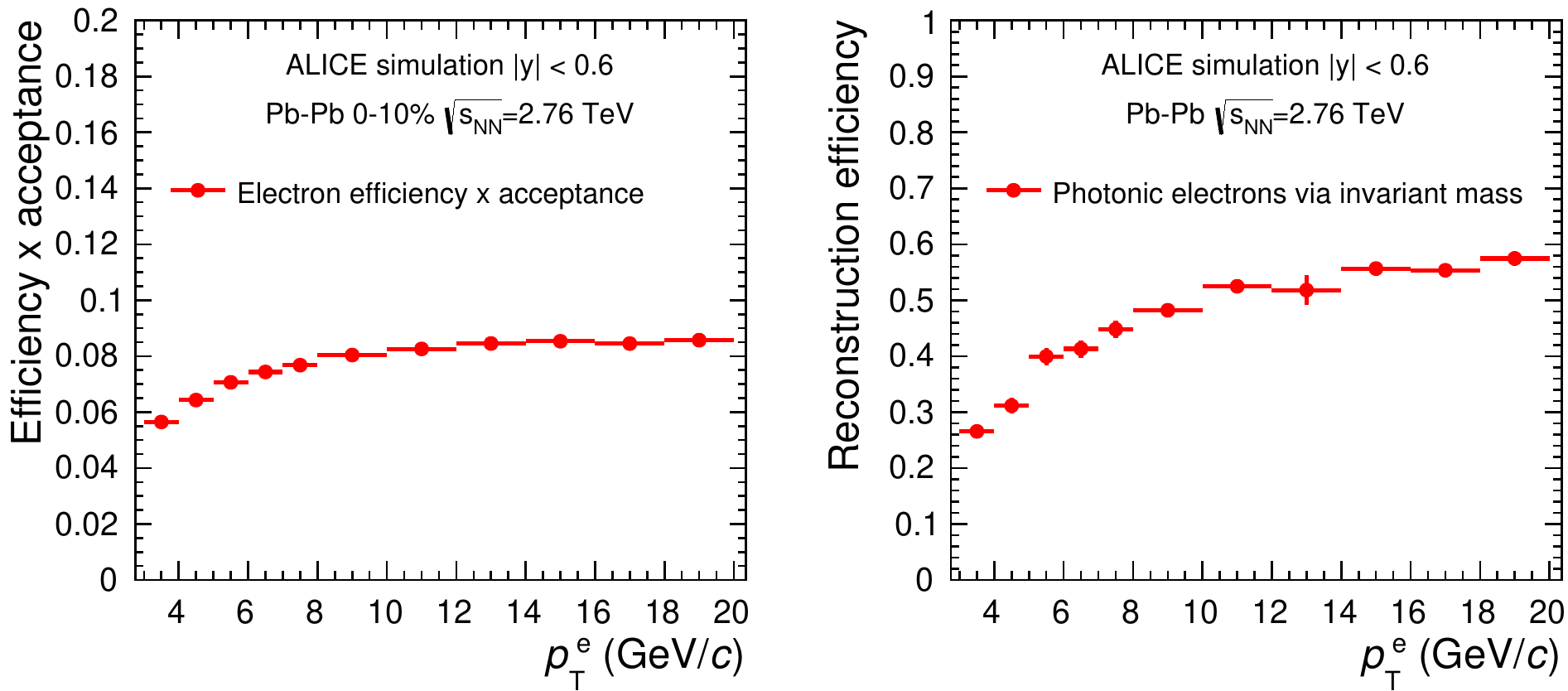}
\end{center}
\caption{{\it \bf Left:} Product of detector acceptance and reconstruction efficiency for inclusive electrons as a function of the electron \pt. The statistical uncertainty is smaller than the size of the points. {\it \bf Right:} Photonic electron reconstruction efficiency via invariant mass ($\epsilon_{e^{\gamma}}$) as a function of \pt\ of the electron.}
\label{fig:electroneffi}
\end{figure}

The efficiencies related to the cuts on the ionisation energy loss in the TPC were estimated with data-driven techniques \cite{Abelev:2014ffa}.
The EMCal efficiencies were calculated using Monte Carlo simulations of proton-proton (PYTHIA \cite{Sjostrand:2006za}) and heavy-ion collisions (HIJING \cite{Wang:1991hta}) with complete detector response modeled by GEANT \cite{Brun:1987ma}.
The product of detector acceptance and reconstruction efficiencies for inclusive electrons for the 10\% most central collisions is shown in the left panel of Fig. \ref{fig:electroneffi}.
The efficiencies were estimated for each centrality class separately.
A variation of about 2.5-3\% was found between the most central (0-10\%) and peripheral (50-80\%) events.

\subsection{Background electron subtraction}

The main sources of electrons contributing to the inclusive electron sample in this analysis are: a) heavy-flavour hadron decay electrons; b) electrons from leptonic decays of quarkonia ($\mathrm{J}/\psi$ and $\Upsilon$ mesons); c) electrons from $W$ and $Z/\gamma^{*}$ decays; d) the so-called {\it photonic} electrons, originating from photon conversions and Dalitz decays of neutral mesons (mainly $\pi^{0}$ and $\eta$); and e) neutral kaon decays; however, the contribution from the non-photonic electrons created in vector meson and $K_{e3}$ decays is negligible ($<0.1\%$) \cite{Abelev:2014gla} in the momentum range considered in this analysis.

The contribution of the photonic electrons to the inclusive electron sample was measured by the invariant mass method.
The invariant mass distribution was determined by pairing every electron track from the inclusive sample with an oppositely-charged track selected with $-3<n_{\sigma}^{\mathrm{TPC}}<$ 3 to increase the chance for finding the pairs.
Pairs satisfying electron identification selections
and pairs satisfying a cut on the invariant mass of $m_{\rm{inv}} < 0.1$ \gevcc\ were selected for further analysis.
These selected unlike-sign pairs, however, contain not only true photonic electrons but also a contribution from random pairs.
This combinatorial background to photonic electrons was estimated using the invariant mass distribution of the like-sign electrons ($N_{\mathrm{e}^{LS}}$), and it was subtracted from that of unlike-sign pairs ($N_{\mathrm{e}^{ULS}}$) to obtain
the number of raw photonic electrons: $N_{\mathrm{e}^{\gamma}}^{raw} = N_{\mathrm{e}^{ULS}} - N_{\mathrm{e}^{LS}}$.

The efficiency for the identification of the photonic electrons by the invariant mass method ($\epsilon_{\mathrm{e}^{\gamma}}$) was estimated from Monte Carlo simulations with full detector response and was found to be centrality independent. The efficiency, shown in the right panel of Fig. \ref{fig:electroneffi}, is about 30\% at $\pt^{e}=4$~\gevc\ and rising to 55\% at 18~\gevc.

The number of photonic electrons present within the inclusive electron sample was calculated as the raw photonic electron yield corrected for the reconstruction efficiency such that: $N_{e^{\gamma}} = N_{e^{\gamma}}^{raw}/\epsilon_{e^{\gamma}}$.
The fraction of photonic electrons within the inclusive electron sample in the 10\% most central collisions is about 30\% at $\pt = 3~\gevc$, drops to 25\% at 12~\gevc\ and remains approximately constant at higher \pT\ considered in this analysis.

The contribution to the inclusive electrons from \jpsi\ decays was estimated using a phenomenological interpolation at $\sqrt{s}$ = 2.76 TeV of the \pt\--differential cross sections measured in \pp\ collisions at various centre-of-mass energies \cite{Bossu:2011qe} and scaling with the nuclear modification factor $R_{\rm{AA}}^{\mathrm{J}/\psi}$ ($p_{\rm{T}}$) measured at the LHC \cite{Adam:2015rba,Abelev:2013ila}.
In 3 $< p_{\mathrm{T}}<$ 4 GeV/$c$, the contribution is 5.5\% in the most central collisions and decreases at high-$p_{\mathrm{T}}$.
The contribution from $\Upsilon$ states estimated from the cross section measured in \pp\ collisions \cite{Chatrchyan:2013yna} was found to be negligible.

The contribution of electrons from W-boson and Z/$\gamma^{*}$ decays was estimated using the cross section obtained from the POWHEG event generator \cite{Oleari:2010nx} for \pp\ collisions and scaled with $\langle T_{\rm{AA}} \rangle$ assuming $R_{\rm{AA}}$ = 1.
The contribution is \pt\ dependent and for W-bosons it increases from 1\% at 10~\gevc\ to about 6\% at 17~\gevc\, whereas the contribution from $Z/\gamma^{*}$ is below 1\% for $\pt<10~\gevc$ and increases to 2.4\% at 17~\gevc.

The heavy-flavour decay electron yield was reconstructed from the inclusive electron yield by first subtracting the photonic electron yield, then correcting the result of the subtraction for the efficiency, and finally, by subtracting the feed-down electrons from $\mathrm{J}/\psi$ and W, Z/$\gamma^{*}$ decays.

\section{Systematic uncertainties}

The sources of systematic uncertainty on the reconstructed heavy-flavour decay electron \pt\ spectrum can be grouped into three categories:
\begin{itemize}
\item event selection (the event normalisation, including the scaling of the EMCal trigger events and the event centrality selection);
\item electron signal extraction (uncertainties originating from corrections related to tracking and particle identification);
\item non-heavy-flavour background determination.
\end{itemize}

An overview of the systematic uncertainties is presented in Table \ref{tab:uncerts}.
For sources that depend on centrality (all but {\it ``tracking/material''} from Table \ref{tab:uncerts}) the uncertainties were evaluated separately in each event class.
In every case a weak centrality dependence was found (deviations of less than 3\%).
In the figures of Section 4 the systematic uncertainties are represented as shaded boxes around the data points.

{\bf Event normalisation.} A comparison of the event normalisation obtained with the EMCal clusters and the normalisation obtained from the inclusive electrons showed a maximum deviation of 8.5\%. This deviation, independent of centrality and \pt, is included as the uncertainty on the yield obtained with the triggered data.
The contribution to the systematic uncertainty due to the 1.1$\%$ relative uncertainty on the fraction of hadronic cross section used in the Glauber fit to determine the centrality is less than 0.1$\%$ in the central event class (0--10$\%$) and 3$\%$ in the semi-peripheral centrality class (50--80$\%$) \cite{Adam:2015nna, ALICE:2012ab}.

{\bf Electron identification.} The systematic uncertainties on the corrections for track reconstruction, track selection and electron identification were assessed via multiple variations of the analysis selections. For each set of cuts the analysis was repeated and compared to the results obtained with the default set of cuts. These variations included changes in track quality cuts, such as the minimum number of the space points in the TPC and associated hits in the ITS. The uncertainties were estimated as a function of track \pt\ and for each centrality class separately. In addition, the electron identification cuts in the TPC ($n_{\sigma}^{\mathrm{TPC}}$) and EMCal ($E/p$ range) were varied around their nominal values. The uncertainty originating from the knowledge of the material budget was estimated via complete detector simulations with the radiation length varied by $\pm 7\%$ \cite{Abelev:2012hxa}.

\begin{table}[t]
   \centering
   \begin{tabular}{@{} lccc @{}} 
      \toprule
      Source                     & \pt\ dependence (GeV/$c$) & Uncertainty (\%) \\
      \midrule
      EMCal trigger correction   & only high-\pt    & 8.5    \\
      Centrality estimation      & n/a              & $<$0.1 - 3    \\
      Tracking / material        & weak within 3-14 & 5  \\
      $E/p$                      & 3 (10)           & 3 (3)  \\
      $n_{\sigma}^{\mathrm{TPC}}$   & 3 (10)            & 3 (7)  \\
      Photonic background        & 3 (10)           & 5 (5)  \\
      \jpsi\ electron background & 3 (10)           & 1 ($<$1) \\
      $W$ electron background		 & 3 (10) 			& 0 ($<$1) \\
      $Z/\gamma^{*}$ electron backgrounds	 & 3 (10)	& $<$1 ($<$2) \\
      \bottomrule
   \end{tabular}
   \caption{Summary of systematic uncertainties on the heavy-flavour electron yields grouped according to their sources. Where applicable the uncertainty was estimated for two \pt\ values, 3 and 10 \gevc\ (for the latter numbers are shown in parentheses) . For details on the extraction of the uncertainties see text.}
   \label{tab:uncerts}
\end{table}

{\bf Subtraction of photonic background.} The uncertainty on the subtracted background electrons from photon conversions and Dalitz decays was obtained by varying the invariant mass cut on the electron pairs within $0.07 < m_{\mathrm{inv}} < 0.15$ GeV/$c^{2}$ and the minimum \pT\ of the tracks paired with electron candidates between 0.3 and 0.6 \gevc.

{\bf Subtraction of electrons from \jpsi .} The uncertainty on the subtracted background electrons from \jpsi\ decays was estimated from the experimental uncertainties on measured production yields in heavy-ion collisions \cite{Adam:2015rba,CMS:2012vxa}.

{\bf Electrons from $W$ and $Z/\gamma^{*}$.} The yield of electrons from $W$ decays was varied by $\pm$~15\% on the basis of the comparison of the $W$ production cross section as given by the POWHEG event generator and the existing measurements in \pp\ collisions at the LHC \cite{Aad:2014qxa}.
The contribution from $Z/\gamma^{*}$ di-electron decays and its uncertainty was estimated using the POWHEG event generator and considered together with the uncertainties on the process production cross section measured in pp collisions \cite{Aad:2014qja}.
Given the small contribution of the electrons from $Z/\gamma^{*}$ decays to the electron spectrum of this analysis the derived uncertainty was found below 1\% at the highest momentum considered.

\section{Results}

\begin{figure}
\begin{center}
\includegraphics[width=1.0\textwidth,viewport=20 24 540 320,clip]{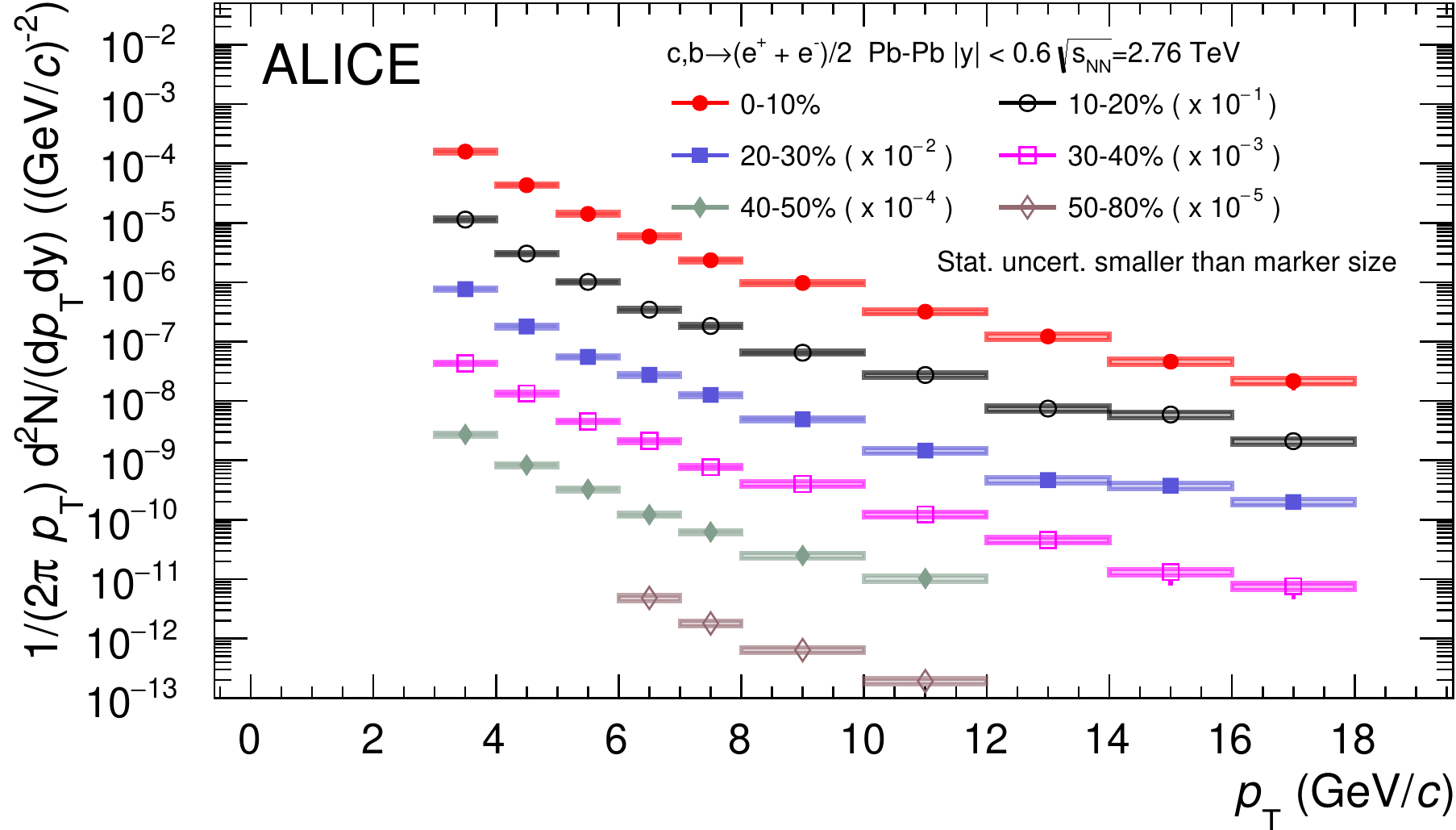}
\end{center}
\caption{Differential yields of electrons from semi-leptonic decays of heavy-flavour hadrons in classes of centrality of \PbPb\ collisions at \sqrtsnn{2.76}.}
\label{fig:hfespectrum}
\end{figure}

The \pt\--differential invariant yields of heavy-flavour decay electrons corrected for acceptance and efficiency in the 0--10\%, 10--20\%, 30--40\%, 40--50\% and 50--80\% centrality classes in Pb--Pb collisions at $\sqrt{s_{\textrm {NN}}}$ =  2.76 TeV are shown in Fig. \ref{fig:hfespectrum}.
Only the EMCal triggered data are shown for the 50-80\% centrality class due to a lack of statistics in the minimum bias data sample.

The production cross section of heavy-flavour decay electrons in \pp\ collisions at \sqrts{2.76}, needed to compute the nuclear modification factor \RAA\ (Eq.~\ref{RAAeq}), was obtained from measurements and FONLL pQCD calculations \cite{Cacciari:2001td,Cacciari:2012ny}.
For $\pt < 12~\gevc$ the measurement at \sqrts{2.76} was used \cite{Abelev:2014gla}.
For $\pt > 12~\gevc$ there is no measurement at this energy.
Thus, an extrapolated cross section was constructed from the measurement at \sqrts{7} by the ATLAS Collaboration \cite{Abelev:2012xe,Aad:2011rr} and the ratio of cross sections at the two collision energies obtained from FONLL \cite{ralf:fonllscale}.
The uncertainties of the pp references are about 20\% for $\pt < 12~\gevc$ and about 15\% for $\pt > 12~\gevc$, including the uncertainty from the scaling with $\sqrt{s}$, which was estimated by consistently varying the FONLL calculation parameters at the two energies \cite{ralf:fonllscale}.

\begin{figure}[t!]
\centering
\includegraphics[width=1.00\textwidth,viewport=5 10 540 370,clip]{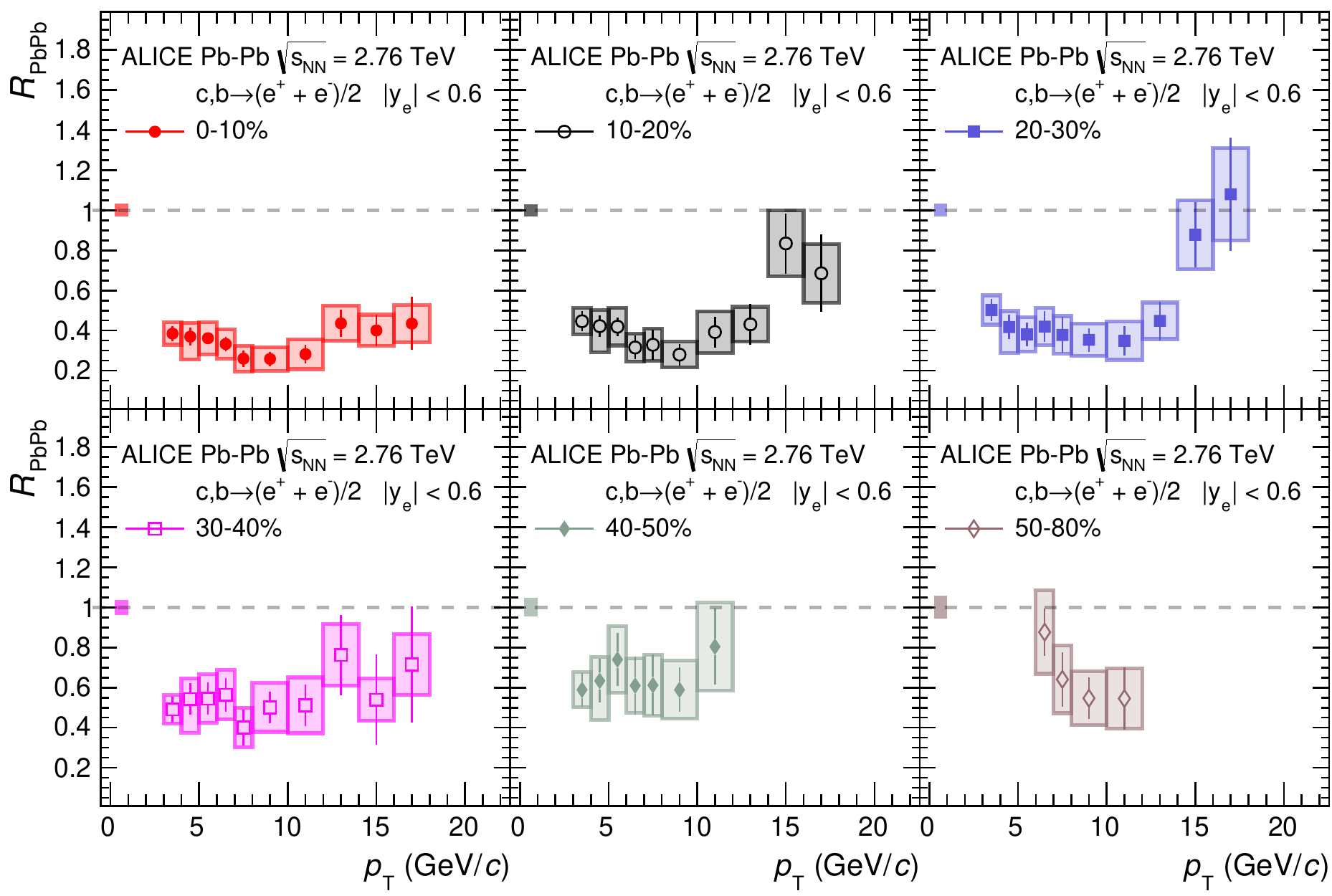}
\caption{\RPBPB\ of electrons from heavy-flavour hadron decays in centrality bins of \PbPb\ collisions at \sqrtsnn{2.76}. The solid band at $\RPBPB=1$ brackets the uncertainty on the average nuclear overlap function ($\langle T_{\rm{AA}} \rangle$).
}
\label{fig:hferaa}
\end{figure}

Figure \ref{fig:hferaa} shows the resulting \RAA\ of heavy-flavour decay electrons for all centrality classes.
The uncertainty on the average nuclear overlap function \TAA\ for each centrality selection was taken as determined in \cite{Abelev:2013qoq}. It varies from 4\% in the 10\% most central events to 7\% in the 50--80\% centrality class, and it is shown as a box at \RAA\ = 1 in the figure.
In all cases, taking into acount the \pt\ trend of the \RAA\ and the statistical uncertainties of the measurement at high-\pt, the electron production yields are suppressed relative to an incoherent superposition of \pp\ collisions.
In the 10\% most central events the \RAA\ reaches values below 0.4, while for the more peripheral events the suppression is weaker.
This centrality dependence of the suppression pattern is qualitatively consistent with in-medium energy loss of heavy quarks due to a decrease of medium's initial energy density and the system size from central to peripheral collisions.

\begin{figure}[t!]
   \centering
   \includegraphics[width=1.0\textwidth,viewport=5 10 560 260,clip]{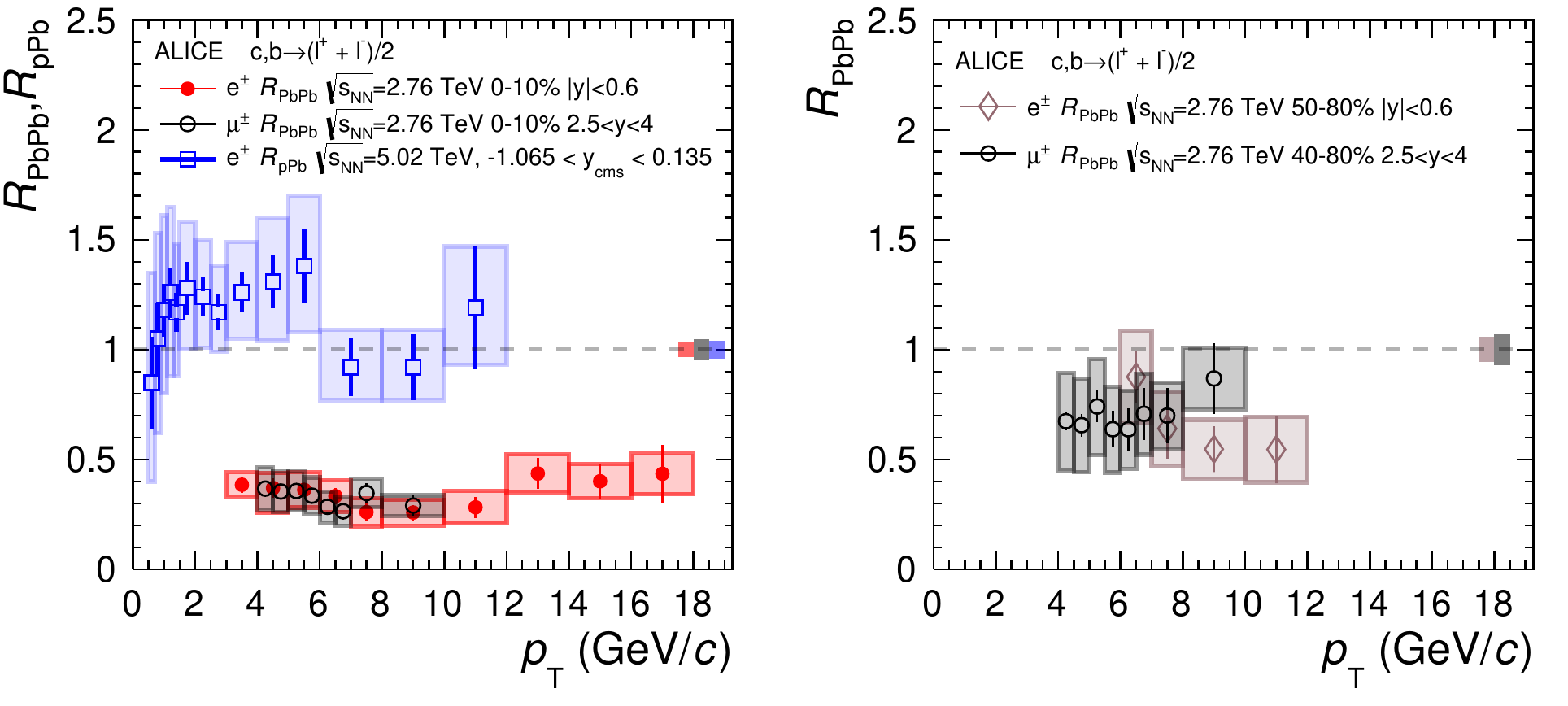}
   \caption{{\it \bf Left:} \RPBPB\ of electrons and muons \cite{Abelev:2012qh} from heavy-flavour hadron decays in 10\% most central \PbPb\ collisions shown together with \RpPb\ of electrons from minimum bias proton-lead collisions at \sqrtsnn{5.02} \cite{Adam:2015qda}. {\it \bf Right:} \RPBPB\ of electrons in semi-peripheral \PbPb\ collisions (50--80\% selection for electrons and 40--80\% for muons.}
   \label{fig:hferaamuons}
\end{figure}

In proton-lead collisions, where formation of a hot, dense and long lived QGP is not expected, the suppression is not observed.
The nuclear modification factor $R_{\textrm{pPb}}$ measured for electrons from heavy-flavour hadron decays is consistent with unity \cite{Adam:2015qda}.
This control measurement in \pPb\ collisions confirms that the strong suppression in \PbPb\ collisions is a result of final state effects.
The left panel of Fig. \ref{fig:hferaamuons} shows the comparison between $R_{\textrm{pPb}}$ for minimum-bias \pPb\ collisions at \sqrtsnn{5.02} and \RAA~for the 10\% most central \PbPb~collisions.
The result reported here for electrons at mid-rapidity is consistent with the measurement of the suppression pattern for muons from the semi-leptonic decays of heavy-flavour hadrons at forward rapidity \cite{Abelev:2012qh}, in both, most central and semi-peripheral collisions (see Fig. \ref{fig:hferaamuons}).
The lepton measurements show remarkable similarity in the suppression pattern that, within the uncertainties, does not exhibit a rapidity dependence.

The \pt\ spectrum of electrons is sensitive to both charm and beauty quark energy loss.
From the decay kinematics and the \pt\--differential cross sections of parent hadrons with charm and beauty, it follows that electrons of \pt\ below 5 GeV/$c$ are mostly sensitive to charm energy loss.
On the other hand, in \pp\ collisions a large fraction (more than 60\%) of the electrons with $\pt>10$~\gevc\ originate from b-quarks \cite{Abelev:2012xe,Abelev:2012sca,Abelev:2014gla,Abelev:2014hla,Cacciari:2012ny}.
The electron yield at high-\pt\ is therefore expected to contain a significant contribution from B mesons with \pt\ up to 30~\gevc.
Consequently, the strong suppression of electrons for $\pt > 10$ \gevc\ is consistent with in-medium energy loss of b-quarks.

\section{Comparison with models}

\begin{figure}[t!]
   \centering
   \includegraphics[width=1.0\textwidth,viewport=10 20 550 430,clip]{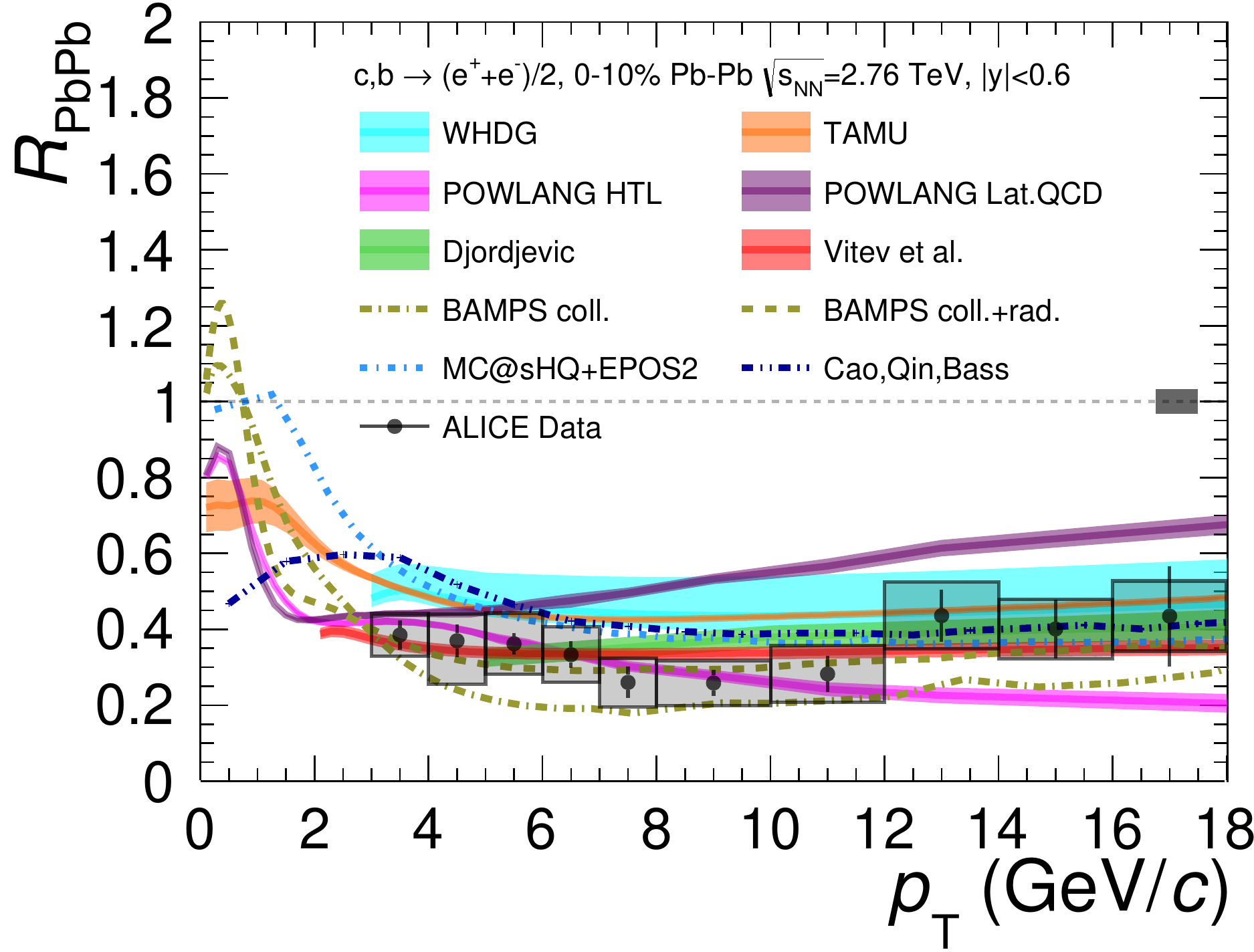}
   \caption{\RPBPB\ of electrons from heavy-flavour hadron decays measured in 10\% most central \PbPb\ collisions at \sqrtsnn{2.76} compared to various theoretical calculations~\cite{Djordjevic:2014tka,Wicks:2005gt,Horowitz:2011gd,Horowitz:2011wm,Sharma:2009hn,He:2014cla,Cao:2013ita,Nahrgang:2013xaa,Alberico:2011zy,Beraudo:2014boa,Uphoff:2011ad,Fochler:2011en,Uphoff:2012gb}.}
   \label{fig:raamodelsNew}
\end{figure}

The \RAA\ of electrons from heavy-flavour hadron decays in the most central Pb--Pb collisions is compared to theoretical models that include heavy quark interactions with the medium in
Fig.~\ref{fig:raamodelsNew}. Most of these models were previously compared to the \RAA\ of D mesons in most central \PbPb\ collisions \cite{ALICE:2012ab,Adam:2015sza} as well as the positive elliptic flow of the D mesons and electrons from heavy-flavour hadron decays in semi-central \PbPb\ collisions \cite{Abelev:2013lca,Abelev:aaa}.
We note that these models differ in the theoretical realisation of the medium properties, and of its dynamics, and also in implementations of the hadronisation and of hadron-hadron interactions in the late stages of the heavy-ion collision.
Also the heavy-quark cross-section used as input to the calculation may differ between the models (PYTHIA, FONLL and POWHEG).

{\bf Djordjevic.} The calculation by {\it Djordjevic} et al. \cite{Djordjevic:2014tka} at $\pt>$~5 GeV/$c$ is consistent with the measurement within the uncertainties including the slow increase of the \RAA\ as a function of electron \pt. The model takes into account both radiative and collisional contributions to parton energy loss. Specifically, the radiative energy loss calculations are an extension of the DGLV \cite{Gyulassy:2000er} model towards a finite size dynamical medium, finite magnetic mass, and running coupling. The model does equally well in reproducing the magnitude and \pt\ dependence of the D mesons \RAA \cite{Adam:2015sza}.

{\bf Vitev.} The calculations by {\it Vitev} et al. \cite{Sharma:2009hn} also capture the magnitude of the suppression and reproduce the \pt\ dependence of the electrons seen in the data.
The in-medium modification of the heavy quark distribution and decay probabilities are evaluated in a co-moving plasma. The predictions for heavy-flavour decay electron suppression are obtained with an improved perturbative QCD description of heavy flavour dynamics in a thermal medium where the formation and dissociation of heavy-flavour mesons are combined with parton-level charm and beauty quark radiative energy loss. The model including the dissociation of heavy-flavour hadrons captures also the suppression of D mesons.

{\bf WHDG.} The band corresponding to the WHDG model calculations \cite{Wicks:2005gt,Horowitz:2011gd,Horowitz:2011wm} is consistent with the measurement within the uncertainties; however, it systematically underpredicts the suppression below 12 \gevc. Interestingly, the same calculation compared to the D mesons \RAA\ reproduced the data very well. The model includes elastic as well as inelastic energy loss of heavy-quarks, and the path length (geometric) fluctuations within a static thermal colored medium with its density as the only free parameter determined via a statistical comparison of the model with the charged particle production in heavy-ion collisions.

{\bf TAMU.} The \RAA\ obtained within the TAMU model of heavy quark transport within a strongly coupled thermal medium including the elastic scatterings with the medium (resonance scattering and coalescence processes) \cite{He:2014cla} underpredicts the suppression at low-\pt\ while it captures the magnitude of the data for $\pt>12~\gevc$.
We note that TAMU also underpredicts the D mesons $R_{\textrm{AA}}$ and its success for the electrons at high-\pt\ may be related to the b-quark energy loss for which the fraction from elastic processes is increased as compared to charm quarks. On the other hand, TAMU reproduces the measured $v_{2}$ of D mesons and electrons from heavy-flavour hadron decays accurately \cite{Abelev:2013lca,Abelev:aaa}.

{\bf BAMPS.} The BAMPS \cite{Uphoff:2011ad,Fochler:2011en,Uphoff:2012gb} calculation, which is a partonic transport model using the Boltzmann equation, is shown for two scenarios.
The {\it BAMPS coll.} calculation considering only the collisional energy loss in an expanding quark-gluon plasma overestimates the magnitude of the suppression within the region covered by the measurement.
The calculation obtained within the same framework where both the elastic and radiative processes were considered ({\it BAMPS coll.+rad.}) describes the data rather well.
A similar conclusion can be drawn from the comparison to the D-meson \RAA.
On the other hand, the {\it BAMPS coll.} reproduces qualitatively the $v_{2}$ of D mesons and electrons from heavy-flavour hadron decays, but the {\it BAMPS coll.+rad.} underestimates the D meson $v_{2}$ \cite{Abelev:2013lca,Abelev:aaa}.

{\bf MC@sHQ+EPOS2.} The results of the Monte Carlo model including a hydrodynamic calculation of the medium coupled with collisional and radiative parton energy loss {\it MC@sHQ+EPOS2} \cite{Nahrgang:2013xaa} are consistent with the measurement within the uncertainties. The model best describes the data at $\pt>12$ GeV/$c$. This model also works better for the D mesons \RAA\ at $\pt>10$~GeV/$c$ as compared to lower momentum (meson \pt\ below 10 GeV/c). The authors of the model emphasize that the scattering in the hadronic phase is not present in their calculation and can have substantial effect on the low-\pt\ suppression and elliptic flow calculations that underpredicts the measurement \cite{Abelev:2013lca,Abelev:aaa}.

{\bf Cao,~Qin,~Bass.} The calculation by {\it Cao, Qin, and Bass} \cite{Cao:2013ita} reproduces the measured \RAA\ at high-\pt\ (above 12 \gevc) while it underpredicts the suppression for low-\pt. The model evaluates the dynamics of energy loss and flow of heavy quarks within the framework of a Langevin equation coupled to a (2+1)-dimensional viscous hydrodynamic model that simulates the space-time evolution of the produced hot and dense QCD matter. This calculation reproduced the suppression of D mesons very accurately, both in strength and the \pt-dependence.

{\bf POWLANG.} The result of the heavy-quark transport calculation using the relativistic Langevin equation with collisional energy loss, POWLANG \cite{Alberico:2011zy,Beraudo:2014boa}, is shown for two choices of heavy-flavour transport coefficients within the quark-gluon plasma. In the {\it POWLANG HTL} \cite{Alberico:2011zy} the coefficients are evaluated by matching the weak-coupling calculations with hard-thermal-loop (HTL) result for soft collisions with a perturbative QCD calculation for hard scatterings. This HTL variant predicts a falling trend with \pt\ of the electrons that is incompatible with the data and overpredicts the suppression at high momentum. Conversely, the calculation that includes the transport coefficients obtained from the Lattice QCD simulations \cite{Beraudo:2014boa} predicts the rising \RAA. However, it reports larger values than the measured ones and it is incompatible with the measured magnitude of the suppression. The width of the theory curves envelopes the spread in the results of the calculation that is obtained when considering two different decoupling temperatures $T_{\textrm {dec}}$ (155 \mev\ and 170 \mev) from the hydrodynamic evolution of the fireball. The relatively small width of the bands suggests a weak sensitivity of the suppression to the $T_{dec}$. Similar to the electron case, POWLANG HTL captures the suppression for D mesons below 5 \gevc\ predicting much lower \RAA\ at high-\pt\ than observed in the data. Interestingly, as in the case of the TAMU model, the POWLANG HTL calculations provide a fair description of the D mesons $v_2$ measured at the LHC.

Given the level of agreement of the theoretical models with the data on $v_{2}$ and \RAA\ of prompt D mesons \cite{ALICE:2012ab,Adam:2015sza,Abelev:2013lca} and electrons from heavy-flavour decays, the following general conclusions arise:
\begin{itemize}
\item models incorporating the complete dynamical and thermal evolution of the medium are favoured by the data;
\item the measurement indicates the need for both, collisional and radiative, energy loss of heavy quarks to be considered to explain the magnitude and the \pt\ dependence of the suppression.
\end{itemize}

\section{Summary}

The \pt--differential yields of electrons from semi-leptonic decays of charm and beauty hadrons were measured at 3$<p_{\textrm T}<$18 GeV/$c$ in several centrality classes of \PbPb\ collisions at \sqrtsnn{2.76} at mid-rapidity.
The nuclear modification factor \RAA\ for the 10\% most central events shows a strong suppression of electrons from heavy-flavour hadron decays.
Consistent with the expectation of a decrease of the medium’s initial energy density and a decreasing system size from central to peripheral collisions, the suppression is significantly weaker in more peripheral events.
No significant suppression is observed in p--Pb collisions, indicating a strong in-medium energy loss of both charm and beauty quarks in Pb--Pb collisions.
In particular, the strong suppression at high-momentum indicates that b-quarks lose a substantial fraction of their energy.
The suppression of electrons is quantitatively consistent with measurements of \RAA\ of muons from semi-leptonic heavy-flavour decays in $2.5 < y < 4$, disfavouring a strong dependence of energy loss on rapidity in the range $|y|<$~4.
Theoretical calculations that include collisional and radiative in-medium energy loss for both charm and beauty quarks reproduce the experimental findings. In particular, models incorporating the dynamical evolution of the medium are preferred by the data.

%
%

\newenvironment{acknowledgement}{\relax}{\relax}
\begin{acknowledgement}
\section*{Acknowledgements}

The ALICE Collaboration would like to thank all its engineers and technicians for their invaluable contributions to the construction of the experiment and the CERN accelerator teams for the outstanding performance of the LHC complex.
The ALICE Collaboration gratefully acknowledges the resources and support provided by all Grid centres and the Worldwide LHC Computing Grid (WLCG) collaboration.
The ALICE Collaboration acknowledges the following funding agencies for their support in building and running the ALICE detector:
A. I. Alikhanyan National Science Laboratory (Yerevan Physics Institute) Foundation (ANSL), State Committee of Science and World Federation of Scientists (WFS), Armenia;
Austrian Academy of Sciences and Nationalstiftung f\"{u}r Forschung, Technologie und Entwicklung, Austria;
, Conselho Nacional de Desenvolvimento Cient\'{\i}fico e Tecnol\'{o}gico (CNPq), Financiadora de Estudos e Projetos (Finep) and Funda\c{c}\~{a}o de Amparo \`{a} Pesquisa do Estado de S\~{a}o Paulo (FAPESP), Brazil;
Ministry of Education of China (MOE of China), Ministry of Science \& Technology of China (MOST of China) and National Natural Science Foundation of China (NSFC), China;
Ministry of Science, Education and Sport and Croatian Science Foundation, Croatia;
Centro de Investigaciones Energ\'{e}ticas, Medioambientales y Tecnol\'{o}gicas (CIEMAT), Cuba;
Ministry of Education, Youth and Sports of the Czech Republic, Czech Republic;
Danish National Research Foundation (DNRF), The Carlsberg Foundation and The Danish Council for Independent Research | Natural Sciences, Denmark;
Helsinki Institute of Physics (HIP), Finland;
Commissariat \`{a} l'Energie Atomique (CEA) and Institut National de Physique Nucl\'{e}aire et de Physique des Particules (IN2P3) and Centre National de la Recherche Scientifique (CNRS), France;
Bundesministerium f\"{u}r Bildung, Wissenschaft, Forschung und Technologie (BMBF) and GSI Helmholtzzentrum f\"{u}r Schwerionenforschung GmbH, Germany;
Ministry of Education, Research and Religious Affairs, Greece;
National Research, Development and Innovation Office, Hungary;
Department of Atomic Energy Government of India (DAE), India;
Indonesian Institute of Science, Indonesia;
Centro Fermi - Museo Storico della Fisica e Centro Studi e Ricerche Enrico Fermi and Istituto Nazionale di Fisica Nucleare (INFN), Italy;
Institute for Innovative Science and Technology , Nagasaki Institute of Applied Science (IIST), Japan Society for the Promotion of Science (JSPS) KAKENHI and Japanese Ministry of Education, Culture, Sports, Science and Technology (MEXT), Japan;
Consejo Nacional de Ciencia (CONACYT) y Tecnolog\'{i}a, through Fondo de Cooperaci\'{o}n Internacional en Ciencia y Tecnolog\'{i}a (FONCICYT) and Direcci\'{o}n General de Asuntos del Personal Academico (DGAPA), Mexico;
Nationaal instituut voor subatomaire fysica (Nikhef), Netherlands;
The Research Council of Norway, Norway;
Commission on Science and Technology for Sustainable Development in the South (COMSATS), Pakistan;
Pontificia Universidad Cat\'{o}lica del Per\'{u}, Peru;
Ministry of Science and Higher Education and National Science Centre, Poland;
Ministry of Education and Scientific Research, Institute of Atomic Physics and Romanian National Agency for Science, Technology and Innovation, Romania;
Joint Institute for Nuclear Research (JINR), Ministry of Education and Science of the Russian Federation and National Research Centre Kurchatov Institute, Russia;
Ministry of Education, Science, Research and Sport of the Slovak Republic, Slovakia;
National Research Foundation of South Africa, South Africa;
Korea Institute of Science and Technology Information and National Research Foundation of Korea (NRF), South Korea;
Centro de Investigaciones Energ\'{e}ticas, Medioambientales y Tecnol\'{o}gicas (CIEMAT) and Ministerio de Ciencia e Innovacion, Spain;
Knut \& Alice Wallenberg Foundation (KAW) and Swedish Research Council (VR), Sweden;
European Organization for Nuclear Research, Switzerland;
National Science and Technology Development Agency (NSDTA), Office of the Higher Education Commission under NRU project of Thailand and Suranaree University of Technology (SUT), Thailand;
Turkish Atomic Energy Agency (TAEK), Turkey;
National Academy of  Sciences of Ukraine, Ukraine;
Science and Technology Facilities Council (STFC), United Kingdom;
National Science Foundation of the United States of America (NSF) and United States Department of Energy, Office of Nuclear Physics (DOE NP), United States.    
\end{acknowledgement}

\bibliographystyle{utphys}   
\bibliography{alicepreprint_CDS}

\newpage
\appendix
\section{The ALICE Collaboration}
\label{app:collab}



\begingroup
\small
\begin{flushleft}
J.~Adam$^\textrm{\scriptsize 39}$,
D.~Adamov\'{a}$^\textrm{\scriptsize 86}$,
M.M.~Aggarwal$^\textrm{\scriptsize 90}$,
G.~Aglieri Rinella$^\textrm{\scriptsize 35}$,
M.~Agnello$^\textrm{\scriptsize 113}$\textsuperscript{,}$^\textrm{\scriptsize 31}$,
N.~Agrawal$^\textrm{\scriptsize 48}$,
Z.~Ahammed$^\textrm{\scriptsize 137}$,
S.~Ahmad$^\textrm{\scriptsize 18}$,
S.U.~Ahn$^\textrm{\scriptsize 70}$,
S.~Aiola$^\textrm{\scriptsize 141}$
% ...
\renewcommand\labelenumi{\textsuperscript{\theenumi}~}

\section*{Affiliation notes}
\renewcommand\theenumi{\roman{enumi}}
\begin{Authlist}
\item \Adef{0}Deceased
\item \Adef{idp1802240}{Also at: Georgia State University, Atlanta, Georgia, United States}
\item \Adef{idp3217328}{Also at: Also at Department of Applied Physics, Aligarh Muslim University, Aligarh, India}
\item \Adef{idp3965744}{Also at: M.V. Lomonosov Moscow State University, D.V. Skobeltsyn Institute of Nuclear, Physics, Moscow, Russia}
\end{Authlist}

\section*{Collaboration Institutes}
\renewcommand\theenumi{\arabic{enumi}~}

$^{1}$A.I. Alikhanyan National Science Laboratory (Yerevan Physics Institute) Foundation, Yerevan, Armenia
\\
$^{2}$Benem\'{e}rita Universidad Aut\'{o}noma de Puebla, Puebla, Mexico
\\
$^{3}$Bogolyubov Institute for Theoretical Physics, Kiev, Ukraine
\\
$^{4}$Bose Institute, Department of Physics 
and Centre for Astroparticle Physics and Space Science (CAPSS), Kolkata, India
\\
$^{5}$Budker Institute for Nuclear Physics, Novosibirsk, Russia
\\
$^{6}$California Polytechnic State University, San Luis Obispo, California, United States
\\
$^{7}$Central China Normal University, Wuhan, China
\\
$^{8}$Centre de Calcul de l'IN2P3, Villeurbanne, Lyon, France
\\
$^{9}$Centro de Aplicaciones Tecnol\'{o}gicas y Desarrollo Nuclear (CEADEN), Havana, Cuba
\\
$^{10}$Centro de Investigaciones Energ\'{e}ticas Medioambientales y Tecnol\'{o}gicas (CIEMAT), Madrid, Spain
\\
$^{11}$Centro de Investigaci\'{o}n y de Estudios Avanzados (CINVESTAV), Mexico City and M\'{e}rida, Mexico
\\
$^{12}$Centro Fermi - Museo Storico della Fisica e Centro Studi e Ricerche ``Enrico Fermi', Rome, Italy
\\
$^{13}$Chicago State University, Chicago, Illinois, United States
\\
$^{14}$China Institute of Atomic Energy, Beijing, China
\\
$^{15}$Commissariat \`{a} l'Energie Atomique, IRFU, Saclay, France
\\
$^{16}$COMSATS Institute of Information Technology (CIIT), Islamabad, Pakistan
\\
$^{17}$Departamento de F\'{\i}sica de Part\'{\i}culas and IGFAE, Universidad de Santiago de Compostela, Santiago de Compostela, Spain
\\
$^{18}$Department of Physics, Aligarh Muslim University, Aligarh, India
\\
$^{19}$Department of Physics, Ohio State University, Columbus, Ohio, United States
\\
$^{20}$Department of Physics, Sejong University, Seoul, South Korea
\\
$^{21}$Department of Physics, University of Oslo, Oslo, Norway
\\
$^{22}$Department of Physics and Technology, University of Bergen, Bergen, Norway
\\
$^{23}$Dipartimento di Fisica dell'Universit\`{a} 'La Sapienza'
and Sezione INFN, Rome, Italy
\\
$^{24}$Dipartimento di Fisica dell'Universit\`{a}
and Sezione INFN, Cagliari, Italy
\\
$^{25}$Dipartimento di Fisica dell'Universit\`{a}
and Sezione INFN, Trieste, Italy
\\
$^{26}$Dipartimento di Fisica dell'Universit\`{a}
and Sezione INFN, Turin, Italy
\\
$^{27}$Dipartimento di Fisica e Astronomia dell'Universit\`{a}
and Sezione INFN, Bologna, Italy
\\
$^{28}$Dipartimento di Fisica e Astronomia dell'Universit\`{a}
and Sezione INFN, Catania, Italy
\\
$^{29}$Dipartimento di Fisica e Astronomia dell'Universit\`{a}
and Sezione INFN, Padova, Italy
\\
$^{30}$Dipartimento di Fisica `E.R.~Caianiello' dell'Universit\`{a}
and Gruppo Collegato INFN, Salerno, Italy
\\
$^{31}$Dipartimento DISAT del Politecnico and Sezione INFN, Turin, Italy
\\
$^{32}$Dipartimento di Scienze e Innovazione Tecnologica dell'Universit\`{a} del Piemonte Orientale and INFN Sezione di Torino, Alessandria, Italy
\\
$^{33}$Dipartimento Interateneo di Fisica `M.~Merlin'
and Sezione INFN, Bari, Italy
\\
$^{34}$Division of Experimental High Energy Physics, University of Lund, Lund, Sweden
\\
$^{35}$European Organization for Nuclear Research (CERN), Geneva, Switzerland
\\
$^{36}$Excellence Cluster Universe, Technische Universit\"{a}t M\"{u}nchen, Munich, Germany
\\
$^{37}$Faculty of Engineering, Bergen University College, Bergen, Norway
\\
$^{38}$Faculty of Mathematics, Physics and Informatics, Comenius University, Bratislava, Slovakia
\\
$^{39}$Faculty of Nuclear Sciences and Physical Engineering, Czech Technical University in Prague, Prague, Czech Republic
\\
$^{40}$Faculty of Science, P.J.~\v{S}af\'{a}rik University, Ko\v{s}ice, Slovakia
\\
$^{41}$Faculty of Technology, Buskerud and Vestfold University College, Tonsberg, Norway
\\
$^{42}$Frankfurt Institute for Advanced Studies, Johann Wolfgang Goethe-Universit\"{a}t Frankfurt, Frankfurt, Germany
\\
$^{43}$Gangneung-Wonju National University, Gangneung, South Korea
\\
$^{44}$Gauhati University, Department of Physics, Guwahati, India
\\
$^{45}$Helmholtz-Institut f\"{u}r Strahlen- und Kernphysik, Rheinische Friedrich-Wilhelms-Universit\"{a}t Bonn, Bonn, Germany
\\
$^{46}$Helsinki Institute of Physics (HIP), Helsinki, Finland
\\
$^{47}$Hiroshima University, Hiroshima, Japan
\\
$^{48}$Indian Institute of Technology Bombay (IIT), Mumbai, India
\\
$^{49}$Indian Institute of Technology Indore, Indore, India
\\
$^{50}$Indonesian Institute of Sciences, Jakarta, Indonesia
\\
$^{51}$Inha University, Incheon, South Korea
\\
$^{52}$Institut de Physique Nucl\'eaire d'Orsay (IPNO), Universit\'e Paris-Sud, CNRS-IN2P3, Orsay, France
\\
$^{53}$Institute for Nuclear Research, Academy of Sciences, Moscow, Russia
\\
$^{54}$Institute for Subatomic Physics of Utrecht University, Utrecht, Netherlands
\\
$^{55}$Institute for Theoretical and Experimental Physics, Moscow, Russia
\\
$^{56}$Institute of Experimental Physics, Slovak Academy of Sciences, Ko\v{s}ice, Slovakia
\\
$^{57}$Institute of Physics, Academy of Sciences of the Czech Republic, Prague, Czech Republic
\\
$^{58}$Institute of Physics, Bhubaneswar, India
\\
$^{59}$Institute of Space Science (ISS), Bucharest, Romania
\\
$^{60}$Institut f\"{u}r Informatik, Johann Wolfgang Goethe-Universit\"{a}t Frankfurt, Frankfurt, Germany
\\
$^{61}$Institut f\"{u}r Kernphysik, Johann Wolfgang Goethe-Universit\"{a}t Frankfurt, Frankfurt, Germany
\\
$^{62}$Institut f\"{u}r Kernphysik, Westf\"{a}lische Wilhelms-Universit\"{a}t M\"{u}nster, M\"{u}nster, Germany
\\
$^{63}$Instituto de Ciencias Nucleares, Universidad Nacional Aut\'{o}noma de M\'{e}xico, Mexico City, Mexico
\\
$^{64}$Instituto de F\'{i}sica, Universidade Federal do Rio Grande do Sul (UFRGS), Porto Alegre, Brazil
\\
$^{65}$Instituto de F\'{\i}sica, Universidad Nacional Aut\'{o}noma de M\'{e}xico, Mexico City, Mexico
\\
$^{66}$Institut Pluridisciplinaire Hubert Curien (IPHC), Universit\'{e} de Strasbourg, CNRS-IN2P3, Strasbourg, France
\\
$^{67}$iThemba LABS, National Research Foundation, Somerset West, South Africa
\\
$^{68}$Joint Institute for Nuclear Research (JINR), Dubna, Russia
\\
$^{69}$Konkuk University, Seoul, South Korea
\\
$^{70}$Korea Institute of Science and Technology Information, Daejeon, South Korea
\\
$^{71}$KTO Karatay University, Konya, Turkey
\\
$^{72}$Laboratoire de Physique Corpusculaire (LPC), Clermont Universit\'{e}, Universit\'{e} Blaise Pascal, CNRS--IN2P3, Clermont-Ferrand, France
\\
$^{73}$Laboratoire de Physique Subatomique et de Cosmologie, Universit\'{e} Grenoble-Alpes, CNRS-IN2P3, Grenoble, France
\\
$^{74}$Laboratori Nazionali di Frascati, INFN, Frascati, Italy
\\
$^{75}$Laboratori Nazionali di Legnaro, INFN, Legnaro, Italy
\\
$^{76}$Lawrence Berkeley National Laboratory, Berkeley, California, United States
\\
$^{77}$Moscow Engineering Physics Institute, Moscow, Russia
\\
$^{78}$Nagasaki Institute of Applied Science, Nagasaki, Japan
\\
$^{79}$National Centre for Nuclear Studies, Warsaw, Poland
\\
$^{80}$National Institute for Physics and Nuclear Engineering, Bucharest, Romania
\\
$^{81}$National Institute of Science Education and Research, Bhubaneswar, India
\\
$^{82}$National Research Centre Kurchatov Institute, Moscow, Russia
\\
$^{83}$Niels Bohr Institute, University of Copenhagen, Copenhagen, Denmark
\\
$^{84}$Nikhef, Nationaal instituut voor subatomaire fysica, Amsterdam, Netherlands
\\
$^{85}$Nuclear Physics Group, STFC Daresbury Laboratory, Daresbury, United Kingdom
\\
$^{86}$Nuclear Physics Institute, Academy of Sciences of the Czech Republic, \v{R}e\v{z} u Prahy, Czech Republic
\\
$^{87}$Oak Ridge National Laboratory, Oak Ridge, Tennessee, United States
\\
$^{88}$Petersburg Nuclear Physics Institute, Gatchina, Russia
\\
$^{89}$Physics Department, Creighton University, Omaha, Nebraska, United States
\\
$^{90}$Physics Department, Panjab University, Chandigarh, India
\\
$^{91}$Physics Department, University of Athens, Athens, Greece
\\
$^{92}$Physics Department, University of Cape Town, Cape Town, South Africa
\\
$^{93}$Physics Department, University of Jammu, Jammu, India
\\
$^{94}$Physics Department, University of Rajasthan, Jaipur, India
\\
$^{95}$Physikalisches Institut, Eberhard Karls Universit\"{a}t T\"{u}bingen, T\"{u}bingen, Germany
\\
$^{96}$Physikalisches Institut, Ruprecht-Karls-Universit\"{a}t Heidelberg, Heidelberg, Germany
\\
$^{97}$Physik Department, Technische Universit\"{a}t M\"{u}nchen, Munich, Germany
\\
$^{98}$Purdue University, West Lafayette, Indiana, United States
\\
$^{99}$Pusan National University, Pusan, South Korea
\\
$^{100}$Research Division and ExtreMe Matter Institute EMMI, GSI Helmholtzzentrum f\"ur Schwerionenforschung, Darmstadt, Germany
\\
$^{101}$Rudjer Bo\v{s}kovi\'{c} Institute, Zagreb, Croatia
\\
$^{102}$Russian Federal Nuclear Center (VNIIEF), Sarov, Russia
\\
$^{103}$Saha Institute of Nuclear Physics, Kolkata, India
\\
$^{104}$School of Physics and Astronomy, University of Birmingham, Birmingham, United Kingdom
\\
$^{105}$Secci\'{o}n F\'{\i}sica, Departamento de Ciencias, Pontificia Universidad Cat\'{o}lica del Per\'{u}, Lima, Peru
\\
$^{106}$Sezione INFN, Bari, Italy
\\
$^{107}$Sezione INFN, Bologna, Italy
\\
$^{108}$Sezione INFN, Cagliari, Italy
\\
$^{109}$Sezione INFN, Catania, Italy
\\
$^{110}$Sezione INFN, Padova, Italy
\\
$^{111}$Sezione INFN, Rome, Italy
\\
$^{112}$Sezione INFN, Trieste, Italy
\\
$^{113}$Sezione INFN, Turin, Italy
\\
$^{114}$SSC IHEP of NRC Kurchatov institute, Protvino, Russia
\\
$^{115}$Stefan Meyer Institut f\"{u}r Subatomare Physik (SMI), Vienna, Austria
\\
$^{116}$SUBATECH, Ecole des Mines de Nantes, Universit\'{e} de Nantes, CNRS-IN2P3, Nantes, France
\\
$^{117}$Suranaree University of Technology, Nakhon Ratchasima, Thailand
\\
$^{118}$Technical University of Ko\v{s}ice, Ko\v{s}ice, Slovakia
\\
$^{119}$Technical University of Split FESB, Split, Croatia
\\
$^{120}$The Henryk Niewodniczanski Institute of Nuclear Physics, Polish Academy of Sciences, Cracow, Poland
\\
$^{121}$The University of Texas at Austin, Physics Department, Austin, Texas, United States
\\
$^{122}$Universidad Aut\'{o}noma de Sinaloa, Culiac\'{a}n, Mexico
\\
$^{123}$Universidade de S\~{a}o Paulo (USP), S\~{a}o Paulo, Brazil
\\
$^{124}$Universidade Estadual de Campinas (UNICAMP), Campinas, Brazil
\\
$^{125}$Universidade Federal do ABC, Santo Andre, Brazil
\\
$^{126}$University of Houston, Houston, Texas, United States
\\
$^{127}$University of Jyv\"{a}skyl\"{a}, Jyv\"{a}skyl\"{a}, Finland
\\
$^{128}$University of Liverpool, Liverpool, United Kingdom
\\
$^{129}$University of Tennessee, Knoxville, Tennessee, United States
\\
$^{130}$University of the Witwatersrand, Johannesburg, South Africa
\\
$^{131}$University of Tokyo, Tokyo, Japan
\\
$^{132}$University of Tsukuba, Tsukuba, Japan
\\
$^{133}$University of Zagreb, Zagreb, Croatia
\\
$^{134}$Universit\'{e} de Lyon, Universit\'{e} Lyon 1, CNRS/IN2P3, IPN-Lyon, Villeurbanne, Lyon, France
\\
$^{135}$Universit\`{a} di Brescia, Brescia, Italy
\\
$^{136}$V.~Fock Institute for Physics, St. Petersburg State University, St. Petersburg, Russia
\\
$^{137}$Variable Energy Cyclotron Centre, Kolkata, India
\\
$^{138}$Warsaw University of Technology, Warsaw, Poland
\\
$^{139}$Wayne State University, Detroit, Michigan, United States
\\
$^{140}$Wigner Research Centre for Physics, Hungarian Academy of Sciences, Budapest, Hungary
\\
$^{141}$Yale University, New Haven, Connecticut, United States
\\
$^{142}$Yonsei University, Seoul, South Korea
\\
$^{143}$Zentrum f\"{u}r Technologietransfer und Telekommunikation (ZTT), Fachhochschule Worms, Worms, Germany
\endgroup

\end{document}